\documentclass[prc,twocolumn,superscriptaddress,showpacs,amssymb,amsmath,amsfonts,aps]{revtex4}
\usepackage{graphicx}
\usepackage{dcolumn}
\begin{document}
\title{Electroexcitation of the $P_{33}(1232)$, $P_{11}(1440)$, $D_{13}(1520)$, 
and $S_{11}(1535)$\\
 at $Q^2=0.4$ and  $0.65~(GeV/c)^2$}

\newcommand*{\YEREVAN }{ Yerevan Physics Institute, 375036 Yerevan, Armenia} 
\affiliation{\YEREVAN } 

\newcommand*{\JLAB }{ Thomas Jefferson National Accelerator Facility, Newport News, Virginia 23606} 
\affiliation{\JLAB } 

\newcommand*{\UCONN }{ University of Connecticut, Storrs, Connecticut 06269} 
\affiliation{\UCONN } 

\newcommand*{\VIRGINIA }{ University of Virginia, Charlottesville, Virginia 22901} 
\affiliation{\VIRGINIA } 

\author{I.G.~Aznauryan}
     \affiliation{\YEREVAN}
\author{V.D.~Burkert}
     \affiliation{\JLAB}
\author{H.~Egiyan}
     \affiliation{\JLAB}
\author{K.~Joo}
     \affiliation{\UCONN}
\author{R.~Minehart}
     \affiliation{\VIRGINIA}
\author{L.C.~Smith}
     \affiliation{\VIRGINIA}
     
\noaffiliation

\begin{abstract}
{Using two approaches: dispersion relations and isobar model,
we have analyzed recent high precision CLAS data
on cross sections of $\pi^0$, $\pi^+$, and $\eta$ electroproduction 
on protons, and the longitudinally
polarized electron beam asymmetry for $p(\vec{e},e'p)\pi^0$ 
and $p(\vec{e},e'n)\pi^+$. The contributions of the
resonances  $P_{33}(1232)$, $P_{11}(1440)$,
$D_{13}(1520)$, $S_{11}(1535)$ to $\pi$ 
electroproduction and
$S_{11}(1535)$ to $\eta$
electroproduction are found. The results
obtained in the two approaches are in good agreement. 
There is also good agreement between 
amplitudes of the $\gamma^* N\rightarrow S_{11}(1535)$
transition found in $\pi$ and $\eta$
electroproduction. For the first time accurate results
are obtained for the longitudinal amplitudes 
of the  $P_{11}(1440)$,
$D_{13}(1520)$, and $S_{11}(1535)$ electroexcitation on protons.
A strong longitudinal response is found for the Roper resonance,
which rules out presentation of this resonance as a hybrid state.}
\end{abstract}

\maketitle

\section{Introduction}
The study of baryon structure is currently in a new 
stage of development as new data with unprecedented precision have become
available due to the high intensity, high duty factor electron accelerators,
complemented with large acceptance multi-particle detectors. 
Very precise measurements of electromagnetically induced hadron production
and its $Q^2$ evolution are available
now, providing a new level in testing of phenomenological
models and QCD in the nonperturbative and subasymptotic domains.
 
Much of our understanding of hadron structure
is connected with the constituent quark model. Simple nonrelativistic
quark models were originally proposed and in many cases
successfully used to explain the relations
between the masses and photocouplings of hadrons.
However, the picture is changing when $Q^2$ grows,
and successes of the simplified quark models do not extend
to the electroexcitation amplitudes.
Once the momentum transfer becomes greater than
the masses of the constituent quarks, a relativistic treatment
of the electromagnetic excitations becomes essential.
The dependence on the small-distance interquark forces
also becomes more significant with increasing $Q^2$.
It is known also that a series of difficulties
in the quark model description of the resonance properties,
such as the unusually slow falloff with  $Q^2$
of the $S_{11}(1535)$ transverse photocoupling amplitude 
and small mass of the
$P_{11}(1440)$,
gave rise to different explanations
of the nature of the nucleon resonances.
Among others, the presence of gluonic degrees of freedom
in the $P_{11}(1440)$ \cite{1} and $K\Sigma$ components in the $S_{11}(1535)$ 
\cite{2,3} have been suggested.
Accurate measurements of the resonance transition
form factors will put all these models to
stringent tests as the internal dynamics of excited states
strongly affects their $Q^2$ evolution.

It is remarkable that this new stage
in the experimental study of the $Q^2$ evolution of
the hadron electromagnetic characteristics  is
paralleled by the significant progress 
in lattice calculations and their 
successes in the description of the 
nucleon and transition form factors \cite{4,5,6}
and nucleon resonances masses \cite{7}. 
Accurate results for the electroexcitation of 
the nucleon resonances will provide important information
for understanding of the $N^*$ structure from the first
principles of QCD by
exploring lattice calculations. 

Recently, precise data on exclusive electroproduction
of $\pi^0$, $\pi^+$, and $\eta$ on protons in the 
first and second resonance  regions
were obtained at Jefferson Lab using the 
CEBAF Large Acceptance Spectrometer
(CLAS). The data include
measurements of the cross sections \cite{8,9,10}, 
including complete angular distributions for the $n\pi^+$ \cite{9} 
and $p\eta$ \cite{10} final states.
The data also include first measurements of the longitudinal
beam asymmetry ($A_{LT'}$) for $p(\vec{e},e'p)\pi^0$ \cite{11}
and $p(\vec{e},e'\pi^+)n$ \cite{12}.  $A_{LT'}$ is very 
sensitive to the interference of transverse and longitudinal amplitudes.
These data allow one to investigate the contributions of resonances
to $\pi$ and $\eta$ electroproduction in greater detail.

The goal of the present investigation is to extract from the data
\cite{8,9,10,11,12} the magnitudes of the
$P_{33}(1232)$, $P_{11}(1440)$,
$D_{13}(1520)$ and $S_{11}(1535)$ resonance contributions 
to these processes.
The analysis is made using two approaches:
(1) fixed-t dispersion relations (DR) and (2) isobar  model (IM), which
were both successfully used in Refs. \cite{13,14} to analyze
$\pi$  and $\eta$ photoproduction.
Comparison of the results obtained in the two
conceptually very different approaches allows us to draw
conclusions on the model dependence of the extracted
resonance characteristics.
The analysis was carried out at $Q^2=0.4$ and  $0.65~(GeV/c)^2$, where 
measurements of $A_{LT'}$ were also made. For these values of $Q^2$ 
we have available the most complete set of CLAS data on  
$\pi^0$, $\pi^+$, and $\eta$ 
cross sections. The formalism is presented
in Section II. In Section III we present
the data and the method of analysis.
In these Sections we also discuss the assumptions
and approximations made in the analysis. 
In Section IV we present and discuss the results, and
a comparison
with previous data and quark model predictions
is made.
Finally, we summarize the results in Section V.

\section{The analysis tools}
The approaches we use to analyze both $\pi$ and $\eta$
electroproduction on protons are dispersion relations 
and the isobar model approximation which are discussed 
in Refs. \cite{13,14} in detail. Here we only briefly present
the main points of these approaches.
\subsection{Dispersion relations}
The imaginary parts of the amplitudes in this approach are built from the 
$s$ channel resonance contributions parametrized in the 
Breit-Wigner form.  All four-, three-, and two-star
resonances from the RPP \cite{15} with masses up to $2.1~GeV$
are taken into account. Using fixed-t dispersion relations 
we find the real parts of the amplitudes. Real parts of 
the amplitudes include
the Born term, i.e. the contributions of the nucleon poles
in the $s$ and $u$ channels, and the contribution
of the $t$ channel $\pi$ exchange in the case of pion electroproduction.
They also include integrals over imaginary parts of the
amplitudes, where we take into account only the contribution
of the resonance energy region. According to estimations
made in Refs. \cite{13,14}, the role of the high energy contributions
to dispersion integrals in the analysis of data in the 
first and second resonance regions is negligibly small.

In the case of pion electroproduction in the elasticity
regions of multipole amplitudes, there is an additional constraint
connected with the Watson theorem. According
to the phase-shift analyses of the $\pi N$ scattering,
the $\pi N$ amplitude $h_{1+}^{3/2}$ is elastic
up to $W=1.45~GeV$ (see, for example, the
GWU(VPI) analysis \cite{16,17}).
By this reason, utilization of the Watson theorem allows us  
to write dispersion relations 
for the amplitudes $M_{1+}^{3/2}$, $E_{1+}^{3/2}$, $S_{1+}^{3/2}$,
which correspond to the $P_{33}(1232)$ resonance,
in the form of integral equations.
In our analysis these amplitudes are presented 
as the solutions of these 
equations. They contain two parts: particular and homogeneous
solutions. Particular solutions have definite magnitudes
fixed by the Born term. Homogeneous solutions have definite shapes
fixed by the integral equations and arbitrary weights
which are fitted parameters in our analysis.

Due to the Watson theorem and large $\pi N$ phases $\delta_{0+}^{1/2}$
and $\delta_{0+}^{3/2}$, a significant contribution to the imaginary
parts of the amplitudes in the  $P_{33}(1232)$ resonance region
can also give nonresonant amplitudes
$E_{0+}^{(0)}$, $E_{0+}^{1/2}$, $E_{0+}^{3/2}$,
$S_{0+}^{(0)}$, $S_{0+}^{1/2}$, and $S_{0+}^{3/2}$.
In order to find these amplitudes we have calculated
their real parts by dispersion relations, taking into
account the Born term and the  $P_{33}(1232)$ resonance
contribution. Then the imaginary parts of the amplitudes were found
using the Watson theorem with the phases  $\delta_{0+}^{1/2}$,
$\delta_{0+}^{3/2}$ taken from the analysis of the 
GWU(VPI) group \cite{16,17}.
Above $W=1.3~GeV$, these contributions were smoothly
reduced to 0.

\subsection{Isobar models}
  
For the isobar model approximation we used 
the approaches developed in Refs. \cite{18,19}
with modifications made in Refs. \cite{13,14}. The main modification
in the case of pion electroproduction is connected with the incorporation
of Regge poles into the unitary isobar model of Ref. \cite{18}.
Due to this modification, the amplitudes of the model transform
with increasing energies into the amplitudes in the Regge pole regime.

Isobar models contain the contributions of resonances
parametrized in Breit-Wigner form, and nonresonant backgrounds
built from the Born terms and the $t$-channel $\rho$ and $\omega$
contributions. In the case of pion production, the background is unitarized
via unitarization of the multipole amplitudes in the $K$-matrix
approximation. Below the two-pion production threshold,
the background multipole amplitudes unitarized in this approximation
satisfy the Watson theorem. At these energies,
the Breit-Wigner formulas for the resonance contributions
are modified in Ref. \cite{13} in such a way that the resonance contributions
to multipole amplitudes also satisfy the Watson theorem.

The Born terms play an important role in both approaches.
They are determined by the coupling constant $g_{\pi NN}$
which is well known, and $g_{\eta NN}$
which we fix according to the results of the analysis of $\eta$
photoproduction data in Ref. \cite{14}.
According to  Ref. \cite{18}, in the unitary isobar model 
that we use for the analysis
of $\pi$ electroproduction, the  $\pi NN$ coupling
being pseudovector at the threshold is transformed with increasing
energy into pseudoscalar coupling. The parameter
that describes this transition was fitted in our analysis;
it turned out to be very close to the value found in  Ref. \cite{18}.

The Born terms are determined also by the nucleon and pion
form factors. We have parametrized these form factors 
in the following way.
According to world data and taking into account
recent measurements (see, for example, review \cite{20}),
we have parametrized proton form factors using the Bosted formulas \cite{21}.
The $Q^2$ evolution of the neutron magnetic form factor
we have described by the dipole formula $G_d(Q^2)=1/(1+Q^2/0.71~GeV^2)^2$,
taking into account the observed deviation from this behaviour \cite{20}.
The value of the neutron electric form factor turned out 
to be very important for the simultaneous description of the $\pi^+$
and $\pi^0$ cross sections in the $P_{33}(1232)$ resonance
region. This form factor was obtained from the requirement
of the best description of these data. As demonstrated below
in Section IVA, the obtained values are in good agreement
with the results of recent measurements \cite{22,23,24,25}.
The pion form factor was parametrized according to recent
measurements \cite{26} by the monopole formula  $1/(1+Q^2/0.54~GeV^2)$.

At $Q^2=0$, the $\rho$ and $\omega$ contributions were found
in the analyses of $\pi$ and $\eta$ photoproduction 
data in Refs. \cite {14,18}. We have introduced the 
 $Q^2$ evolution of these contributions via the description
of the $\rho(\omega)\rightarrow \pi \gamma$ and
$\rho(\omega)\rightarrow \eta \gamma$ form factors
using the dipole formula $G_d(Q^2)$.
In order to take into account the uncertainty of these 
form factors, the $\rho$ and $\omega$ contributions
were allowed to vary in the vicinity of their values
found in the above described way.
It turned out that the obtained deviations
do not exceed $20\%$.  

\begin{figure}[t]
\includegraphics[scale=0.45]{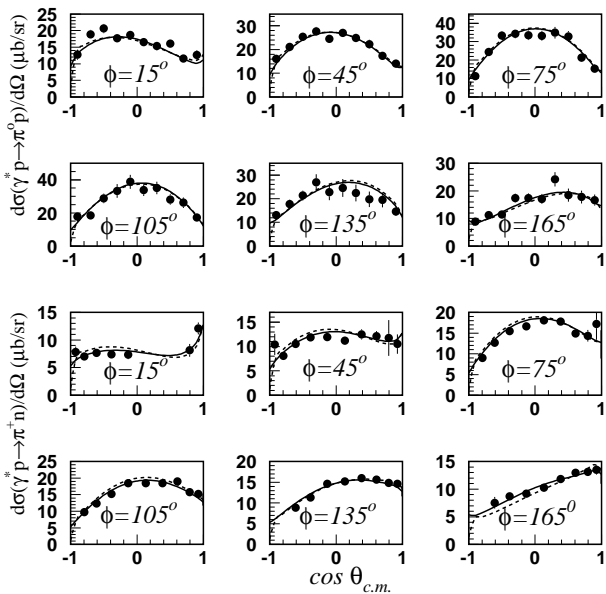}
\caption{ Differential cross sections at $Q^2=0.4~(GeV/c)^2$:
$W=1.22~GeV$ for $\gamma^* p\rightarrow \pi^0 p$
and  $W=1.23~GeV$ for $\gamma^* p\rightarrow \pi^+ n$.
The solid and dashed curves are the results obtained using
IM and DR, respectively.
The data are from CLAS \cite{8,9}.}
\label{fig:fig1}
\end{figure}

\section{Data and Analysis}

We have analyzed the following sets of data at  
$Q^2=0.4$ and  $0.65~(GeV/c)^2$.

(a) At $Q^2=0.4~(GeV/c)^2$ the data used in our analysis
are the results of recent CLAS measurements of 
$\pi^0$ ($W=1.1-1.68~GeV$) \cite{8}   
and $\pi^+$ ($W=1.1-1.55~GeV$) \cite{9}
differential cross sections, and polarized beam asymmetry
in $\pi^0$ 
and $\pi^+$ electroproduction ($W=1.1-1.66~GeV$) \cite{11,12}.    

(b) At $Q^2=0.65~(GeV/c)^2$ we have used recent CLAS measurements
of  $\pi^0$ electroproduction cross sections 
($W=1.1-1.52~GeV$, $E_e=1.645~GeV$ and 
$W=1.1-1.68~GeV$, $E_e=2.445~GeV$) \cite{8},
and polarized beam asymmetry
in $\pi^0$
and $\pi^+$ electroproduction ($W=1.1-1.66~GeV$) \cite{11,12}.
We have also used CLAS data on  $\pi^+$ 
differential cross sections ($W=1.1-1.41~GeV$, $Q^2=0.6~(GeV/c)^2$) \cite{9}.
As the values of $Q^2$ in \cite{9}
and the main data set are different, and the data 
on $\pi^+$ differential cross sections 
extend over more restricted range in $W$, we have complemented this
data set by the results of DESY for $\pi^0$    
and $\pi^+$ differential cross sections
at $W=1.415-1.505~GeV$, $Q^2=0.6-0.63~(GeV/c)^2$ \cite{27,28}
and $W=1.565-1.655~GeV$, $Q^2=0.6-0.64~(GeV/c)^2$ \cite{29,30}, and
by the results of NINA for $\pi^0$                            
differential cross sections
at $W=1.395-1.425~GeV$, $Q^2=0.6,~0.61~(GeV/c)^2$ \cite{31}.

\begin{figure}[t]
\includegraphics[scale=0.45]{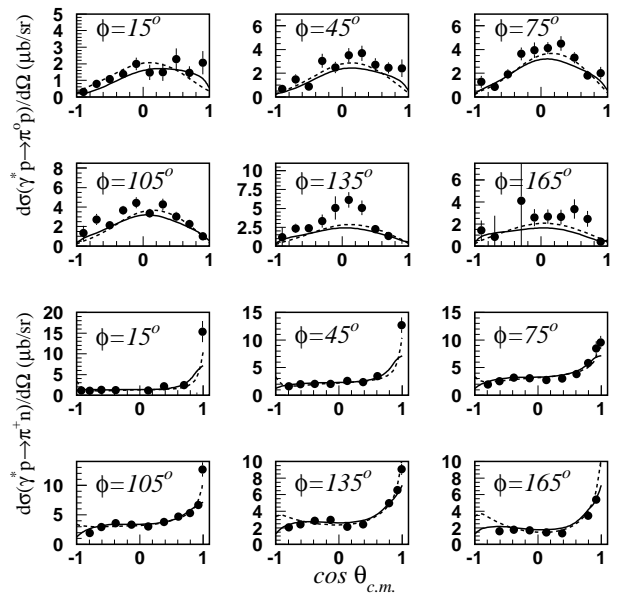}
\caption{Differential cross sections at $Q^2=0.4~(GeV/c)^2$:
$W=1.52~GeV$ for $\gamma^* p\rightarrow \pi^0 p$  
and  $W=1.53~GeV$ for $\gamma^* p\rightarrow \pi^+ n$.
The solid and dashed curves are the results obtained using
IM and DR, respectively.
The data are from CLAS \cite{8,9}.}
\label{fig:fig2}
\end{figure}

\begin{figure}[h]
\includegraphics[scale=0.45]{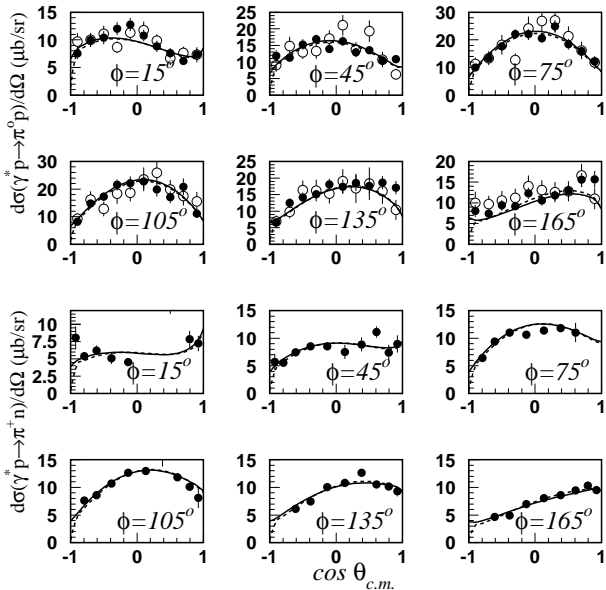}
\caption{Differential cross sections  
for $\gamma^* p\rightarrow \pi^0 p$
at $Q^2=0.65~(GeV/c)^2$ and $W=1.22~GeV$ 
and $\gamma^* p\rightarrow \pi^+ n$
at $Q^2=0.6~(GeV/c)^2$ and $W=1.23~GeV$.
The solid and dashed curves are the results obtained using
IM and DR, respectively.
The data are from CLAS \cite{8,9}.
For $\gamma^* p\rightarrow \pi^0 p$, open and solid circles correspond
to measurements with  $E_e=1.645$ and  $2.445~GeV$ \cite{8}, respectively. }
\label{fig:fig3}
\end{figure}

Let us note that unlike old measurements, which extend mostly
over limited ranges of angles, the CLAS data cover the full angular range.

Both data sets include first, second, and partly third resonance
regions. The full angular coverage of the CLAS data, and the presence
of the $A_{LT'}$ data, which are sensitive both to transverse
and longitudinal amplitudes,
allowed us to reliably evaluate the resonance
contributions in the first and second resonance
regions. 

The analysis of old data 
has been performed in Ref. \cite{32} 
with inclusion from data sets (a) and (b)
only for $\pi^0$ differential cross sections. 
The obtained results qualitatively agree with our's,
however, there are significant differences in details,
and in Ref. \cite{32} the longitudinal helicity amplitude 
for the $\gamma^* p\rightarrow D_{13}(1520)$ transition
is not found. As shown below, in our analysis
this amplitude is found with good accuracy.

The fitted parameters in our analysis were the magnitudes
of the multipole amplitudes corresponding to the 
contributions of the resonances 
from the first and second resonance
regions at resonance positions. 
The magnitudes of the amplitudes
corresponding to the most prominent resonance
of the third resonance region, the $F_{15}(1680)$, were also fitted.
The transverse amplitudes of the $S_{31}(1620)$, $S_{11}(1650)$, $D_{15}(1675)$,
$D_{13}(1700)$ and $D_{33}(1700)$ were fixed according to
the results of the analysis made within the single quark
transition model in Ref. \cite{33}. The longitudinal
amplitudes of these resonances and the amplitudes
of other resonances were fixed taking into account the results
obtained in the analysis in Ref. \cite{34}. In order
to check the stability of our results,
we also performed an analysis where the amplitudes
of all resonances from the third resonance region
were also fitted. The obtained 
results for the amplitudes in the 
first and second resonance
regions remained stable.

\begin{figure}[h]
\includegraphics[scale=0.45]{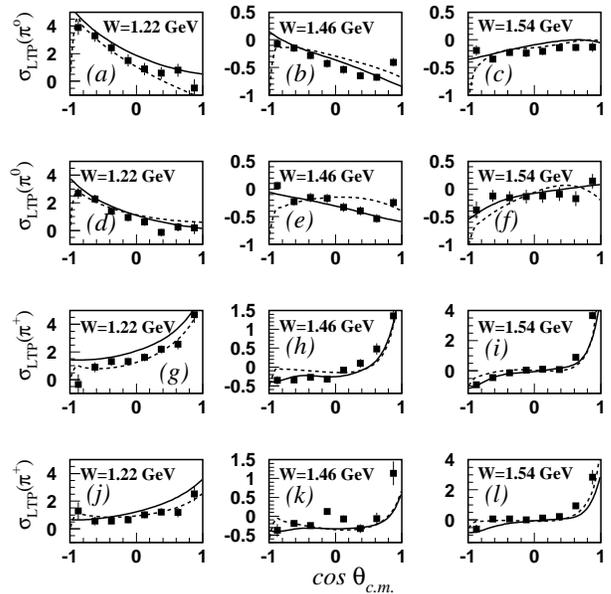}
\caption{Structure function $\sigma_{LT'}$  
for $\gamma^* p\rightarrow \pi^0 p$
at $Q^2=0.4~(GeV/c)^2$ (Figs. a-c) and $Q^2=0.65~(GeV/c)^2$ (Figs. d-f),
and for $\gamma^* p\rightarrow \pi^+ n$
at $Q^2=0.4~(GeV/c)^2$ (Figs. g-i) and $Q^2=0.65~(GeV/c)^2$ (Figs. j-l).
The solid and dashed curves are the results obtained using
IM and DR, respectively.
The data are from CLAS \cite{11,12}.}
\label{fig:fig4}
\end{figure}

\begin{figure}[h]
\includegraphics[scale=0.45]{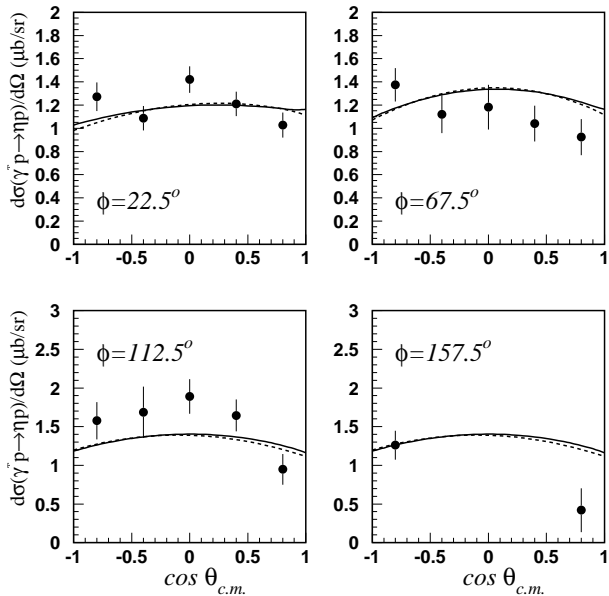}
\caption{Differential cross sections 
for $\gamma^* p\rightarrow \eta p$
at $Q^2=0.375~(GeV/c)^2$
and $W=1.53~GeV$.
The solid and dashed curves are the results obtained using
IM and DR, respectively.
The data are from CLAS \cite{10}.}
\label{fig:fig5}
\end{figure}

It is known that  $\eta$ photo- and electroproduction
provide a unique opportunity to study the $S_{11}(1535)$
resonance, because the contributions of nearby resonances
are strongly suppressed in comparison with the $S_{11}(1535)$ contribution. 
In order to obtain additional information
on the electroexcitation of the $S_{11}(1535)$,
we have also performed an analysis
of the CLAS data on $\eta$ electroproduction cross sections
at $Q^2=0.375~(GeV/c)^2$ ($W=1.5-1.62~GeV$)
and $Q^2=0.75~(GeV/c)^2$ ($W=1.5-1.83~GeV$) \cite{10}.
In our analysis of these data we have taken into
account the results obtained in the analyses of
$\eta$ photoproduction in Ref. \cite{14}  and of
$\pi$  electroproduction in this work.
The contributions of all resonances, except
the $S_{11}(1535)$, were fixed according to these results.
With this, the $\eta N$ branching ratios for
the $D_{13}(1520)$ and  $F_{15}(1680)$ were taken from Ref. \cite{14}.
Other $\eta N$ branching ratios and the branching ratios
to $\pi N$ channel were taken from the RPP \cite{15}.
So, only the amplitudes of the $S_{11}(1535)$
were fitted in our analysis of $\eta$ data.

\begin{table}
\begin{center}
\begin{tabular}{cccccc}
\hline
${}$&${}$&Number&${}$&$\chi2/data$&${}$\\
Observable&$Q^2$&of&${}$&${}$&${}$\\
${}$&${}$&data~points&$IM$&${}$&$DR$\\
\hline
$\frac{d\sigma}{d\Omega}(\pi^0)$&0.4&3530&1.22&&1.21\\
&0.6-0.65&6537&1.22&&1.39\\
\hline
$\frac{d\sigma}{d\Omega}(\pi^+)$&0.4&2308&1.62&&1.97\\
&0.6-0.65&1716&1.48&&1.75\\
\hline
$A_{LT'}(\pi^0)$&0.4&956&1.14&&1.25\\
&0.65&805&1.07&&1.3\\
\hline
$A_{LT'}(\pi^+)$&0.4&918&1.18&&1.63\\
&0.65&812&1.18&&1.15\\
\hline
$\frac{d\sigma}{d\Omega}(\eta)$&0.375&172&1.32&&1.33\\
&0.75&412&1.42&&1.45\\
\hline
\end{tabular}
\caption{\label{tab1} Obtained values of $\chi2$.}
\end{center}
\end{table}

\section{Results and Discussion}
The results of our fit to the data are presented in Table I.
The description is good for all
observables in both approaches. The somewhat larger value
of $\chi^2$ for the $\pi^+$ cross sections
is associated with the small statistical uncertainties of these
measurements \cite{9}.
In Figs. 1-6 we present our results for observables
in comparison with experimental data.

\begin{figure}[h]
\includegraphics[scale=0.45]{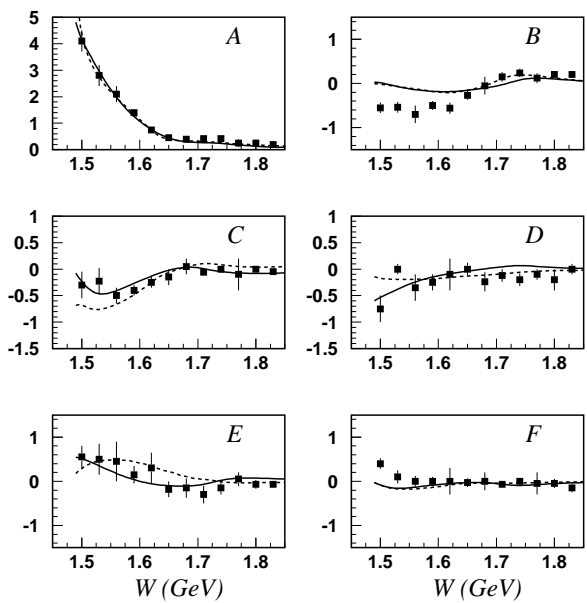}
\caption{Coefficients $A,~B,~C,~D,~E,~F$   at $Q^2=0.75~(GeV/c)^2$
in the expansion of 
$\gamma^* p\rightarrow \eta p$ 
differential cross section over 
$cos\theta$ \cite{10}.
The solid and dashed curves are the results obtained using
IM and DR, respectively.
The data are from CLAS \cite{10}.}
\label{fig:fig6}
\end{figure}

\begin{figure}[h]
\includegraphics[scale=0.45]{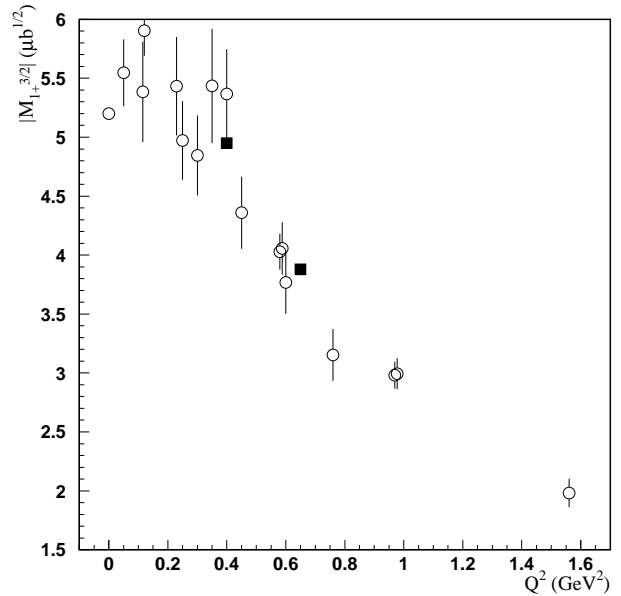}
\caption{The amplitude $|M_{1+}^{3/2}|$ at $W=1.229~GeV$, where according
to the GWU(VPI) analysis of pion 
photoproduction \cite{17,35}  $ReM_{1+}^{3/2}=0$.
Full boxes are the average values of our results obtained using
IM and DR.
Open circles correspond to Refs. \cite{36,37,38,39,40}.
Open circle at $Q^2=0$ is the result of the GWU(VPI) 
analysis of pion photoproduction 
data \cite{17,35}. }
\label{fig:fig7}
\end{figure}

\subsection{$P_{33}(1232)$}
The $P_{33}(1232)$ is an isolated resonance located
in the region where the corresponding $\pi N$ amplitude is elastic.
For this reason, the $P_{33}(1232)$ is investigated with great accuracy
in $\pi N$ scattering and in $\pi$ photoproduction.
Due to isotopic invariance violation and possible presence
of nonresonant contributions, there is a difference
between the phases $\delta_{1+}^{3/2}$ found in these reactions.
According to the analyses of the GWU(VPI) group \cite{16,17,35},
the value of $W$, where $\delta_{1+}^{3/2}=\pi/2$, is
$\simeq 1.232~GeV$ in $\pi N$ scattering,
and $\simeq 1.229~GeV$ in $\pi$ photoproduction. 
In our analysis we have used the phase $\delta_{1+}^{3/2}$
found in the GWU(VPI) analysis of pion photoproduction, and the amplitudes
presented in Table II correspond to $W=1.229~GeV$.
We have checked the possibility of shifting this phase;
however, in both approaches this was not supported by the fit
to the data. 

From Table II it can be seen that the amplitudes obtained in the two analyses
are in good agreement with each other. The values
of the ratios $ImE_{1+}^{3/2}/ImM_{1+}^{3/2}$, 
$ImS_{1+}^{3/2}/ImM_{1+}^{3/2}$ are well determined,
and are in good agreement with the results
of Ref. \cite{8} obtained from $\pi^0$ cross sections 
using a multipole expansion. 

In Fig. 7 we compare our results for  $M_{1+}^{3/2}$
with the values obtained in coincidence
$\pi^0$ electroproduction experiments \cite{36,37,38,39,40}
using multipole analysis.
Significantly smaller uncertainties are obtained 
from the new CLAS data.

There are also the values of $M_{1+}^{3/2}$
obtained in inclusive experiments \cite{41,42,43}.
Our analysis shows that these values are in very good
agreement with our results. However, the $M_{1+}^{3/2}$ amplitudes 
obtained from inclusive data are not very reliable,
as no partial-wave analysis is possible in this case, and 
the separation of the transverse
and longitudinal contributions in these measurements
is arbitrary. For this reason we do not present here
the comparison with the results obtained in inclusive
experiments.

It is interesting to note that the possibility of 
simultaneously describing the  $\pi^0$ and $\pi^+$ electroproduction
cross sections
in the $P_{33}(1232)$ resonance region depends
significantly on the value of the neutron electric
form factor $G_{en}(Q^2)$, which enters $\gamma^* p\rightarrow \pi^+ n$
through the $u$ channel neutron exchange.
This form factor gives significant contribution to the multipole
$E_{0+}(\gamma^* p\rightarrow \pi^+ n)$,
which has a large contribution to $\pi^+$ electroproduction
cross section at $W<1.3~GeV$. For example,
$G_{en}(Q^2)=0.06$ reduces $E_{0+}(\gamma^* p\rightarrow \pi^+ n)$
at $Q^2=0.4$ and $W=1.23~GeV$ by $13~\%$. By this reason $G_{en}(Q^2)$
significantly affects the value of the resonance multipole
amplitude $M_{1+}^{3/2}$ which is needed to describe
 $\pi^+$ electroproduction
cross section in the $P_{33}(1232)$ resonance region.
In both our approaches we have found that a 
successful simultaneous description of the
$\pi^0$ and $\pi^+$ electroproduction
cross sections requires values of
$G_{en}(Q^2)=0.06$ and $0.05$ at $Q^2=0.4$ and $0.65~(GeV/c)^2$,
respectively. These values are in good agreement
with the results of recent measurements: 
$G_{en}(Q^2)=0.052\pm 0.004$ \cite{22} at $Q^2=0.4~(GeV/c)^2$,  
$G_{en}(Q^2)=0.046\pm 0.006\pm 0.003$ \cite{23},
$G_{en}(Q^2)=0.053\pm 0.003\pm 0.003$ \cite{25}
at $Q^2=0.5~(GeV/c)^2$, and  
$G_{en}(Q^2)=0.048\pm 0.007$ \cite{24} at $Q^2=0.67~(GeV/c)^2$.  
\subsection{$P_{11}(1440),~D_{13}(1520),~S_{11}(1535)$}
The results for these resonances are presented
in Tables III and IV. The quoted errors are  the uncertainties
arising from the fitting procedure.
Comparison of the results obtained in the two
approaches allows us to estimate the model
dependence of the extracted  amplitudes.
It is seen that the two approaches give very close results,
except for the transverse helicity amplitude for the
$P_{11}(1440)$ electroexcitation where the difference
of the results obtained using the two approaches is
significantly larger than the fit uncertainties. Nevertheless,
we can make definite conclusion that the $A_{1/2}$
amplitude of the $\gamma^*p\rightarrow
P_{11}(1440)$ transition falls very rapidly
in comparison with its value at $Q^2=0$.

In Figs. 8-9 we present our results for the helicity
amplitudes of the $\gamma^*p\rightarrow 
P_{11}(1440)$, $D_{13}(1520)$, and $S_{11}(1535)$
transitions along with the results at 
the photon point from the RPP \cite{15},
where the outcomes of various analyses are combined.
Separately, at $Q^2=0$, we present the results
for the $S_{11}(1535)$ found in $\pi$ \cite{16} and 
$\eta$ \cite{14} photoproduction.
At $Q^2>0$, data on helicity amplitudes are more sparse
and available only for the $D_{13}(1520)$, $S_{11}(1535)$
transverse amplitudes. These data are analyzed
in Ref. \cite{33} on the basis of the single quark
transition model. As a result, bands that correspond
to the existing data from Bonn, DESY, and NINA, and JLab
measurements of $\eta$ electroproduction \cite{10} are found.
These bands are presented in Figs. 8-9 in the form
of shadowed areas. 

From Figs. 8-9 it is seen that our results
for the $D_{13}(1520)$ and $S_{11}(1535)$
transverse amplitudes are in good agreement with the results
that follow from the previous data.
It should be noted, however, that our results
give a much more definite value for the ${}_pA_{1/2}$
amplitude of the $\gamma^*p\rightarrow D_{13}(1520)$
transition than earlier results.
The transverse amplitude ${}_pA_{3/2}$ for this transition, 
which is dominant
at $Q^2=0$, falls very rapidly with increasing  $Q^2$,
and all quark models predict that
at high $Q^2$,  ${}_pA_{1/2}$ will become the 
dominant contribution to $\gamma^*p\rightarrow D_{13}(1520)$. 
Such behavior is confirmed by our analysis of the CLAS data,
with the crossover where ${}_pA_{1/2}={}_pA_{3/2}$ occurring
around $Q^2 = 0.4-0.65~(GeV/c)^2$. 

From Fig. 8 it is seen that the transverse helicity amplitude
${}_pA_{1/2}$ of the $P_{11}(1440)$ shows a rapid fall off with $Q^2$.
Moreover,
at $Q^2\simeq 0.5-0.6~(GeV/c)^2$ this amplitude, apparently, changes
sign. 

Amplitudes corresponding to the $S_{11}(1535)$
have been investigated in our analysis of $\pi$ and $\eta$ electroproduction.
For consistent comparison of these amplitudes
and for comparison with the results obtained
in other works, following  Ref. \cite{10},
we have used the parameters: $M=1530~MeV$,
$\Gamma_{tot}=150~MeV$, $\beta_{\pi N}=0.4$ and $\beta_{\eta N}=0.55$.
It is seen that unlike at $Q^2=0$,
where the amplitude ${}_pA_{1/2}$ found in $\pi$ photoproduction
is much smaller than the amplitude
found in $\eta$ photoproduction, 
our results for $\pi$ and $\eta$ electroproduction
are in good agreement with each other.

In Figs. 8-9 we compare our results with quark model
predictions. As mentioned in the Introduction,
with increasing $Q^2$ a relativistic treatment of the electroexcitations
of the nucleon resonances becomes essential. The most appropriate
way to realize this is to consider
electromagnetic transition amplitudes in the light-front
dynamics (see, for example, Refs. \cite{44,45,46}). 
Such an approach has been used to investigate 
the $Q^2$ evolution  of the $\gamma^*p\rightarrow
P_{11}(1440)$, $D_{13}(1520)$, and $S_{11}(1535)$
transitions in Ref. \cite{46}. From the results of this investigation 
the important role of relativistic effects is quite evident:
bold and thin solid curves in Figs. 8-9 correspond
to relativistic and nonrelativistic calculations.
Light-front dynamics has also been used in Refs. \cite{47,48}
for investigation of the  $\gamma^*p\rightarrow
P_{11}(1440)$, $S_{11}(1535)$ transition amplitudes.
Different forms of interquark forces inspired by QCD
are considered in Refs. \cite{49,50}, where the relativistic effects
are taken into account only partly. 
From Figs. 8-9  it can be seen that none of 
the approaches \cite{46,47,48,49,50} gives simultaneously a good description
of the $\gamma^*p\rightarrow
P_{11}(1440)$, $D_{13}(1520)$, $S_{11}(1535)$
transition amplitudes.

In Ref. \cite{51} for the $\gamma^*p\rightarrow
P_{11}(1440)$ transition, the contribution of the diagram 
where the photon interacts
with the $q\bar{q}$ cloud of the nucleon
is taken into account in addition 
to the main mechanism where electroexcitation
of the nucleon occurs via photon interaction with
constituent quarks. From Fig. 8
it can be seen that the predictions of Ref. \cite{51}
are qualitatively similar to the results of the light-front approach
\cite{46} and are in good agreement with our results.
However, for definite conclusions it is necessary to also
have predictions of the Ref. \cite{51} approach for
the $\gamma^*p\rightarrow D_{13}(1520)$, $S_{11}(1535)$
transitions. 

In Fig. 8 we have also presented the predictions for 
the $P_{11}(1440)$
obtained by assuming that the $P_{11}(1440)$ 
is a $q^3 G$ hybrid state \cite{1}.
It is seen that the longitudinal helicity amplitude
of the $\gamma^*p\rightarrow
P_{11}(1440)$ transition obtained under this assumption
strongly contradicts our results.

\begin{figure}[t]
\includegraphics[scale=0.45]{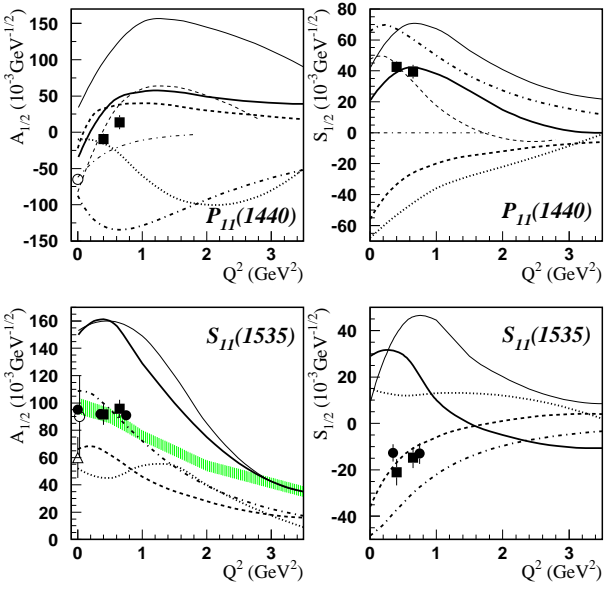}
\caption{Helicity amplitudes for the $P_{11}(1440)$ 
and $S_{11}(1535)$ electroexcitations on protons.
Full  boxes are the average values of our results obtained 
in the analysis of $\pi$ electroproduction data using
IM and DR.
Full circles are the average values of the results obtained 
using IM and DR
in the analysis of $\eta$ photo and electroproduction data in
Ref. \cite{14} and in this work. 
Open triangle at $Q^2=0$ is the result of 
the GWU(VPI) analysis of pion photoproduction
data \cite{16}.
Open circles at $Q^2=0$ are the RPP estimates \cite{15}.
Shadowed area
\cite{33} corresponds to the results obtained
from the existing Bonn, DESY, and NINA data, and from JLab
measurements of $\eta$ electroproduction \cite{10}.
Bold and  thin solid curves  correspond
to relativistic and nonrelativistic 
quark model calculations in Ref. \cite{46}.
Bold dashed curves correspond
to the light-front calculations of  Refs. \cite{47,48}.
Dotted, bold dashed-dotted and thin dashed curves correspond
to the quark models of  Refs. \cite{49,50,51}.
Thin dashed-dotted curves are the predictions 
obtained assuming that the $P_{11}(1440)$
is a $q^3 G$ hybrid state \cite{1}.}
\label{fig:fig8}
\end{figure}

\begin{figure}[t]
\includegraphics[scale=0.45]{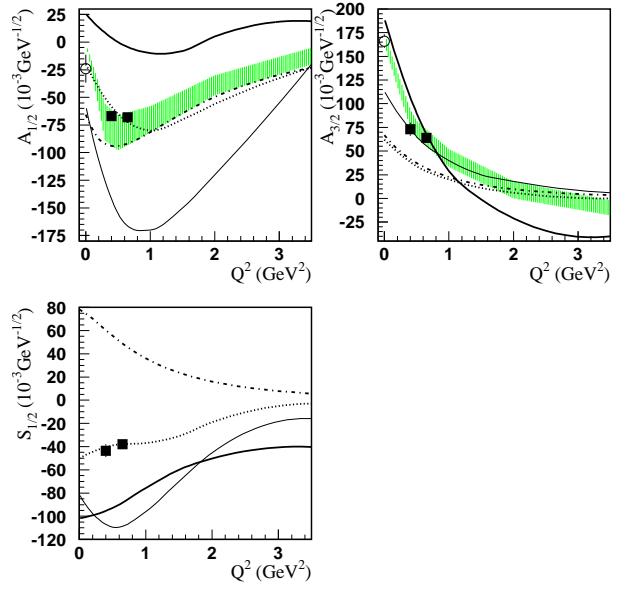}
\caption{Helicity amplitudes for the $D_{13}(1520)$ electroexcitation.
All other relevant information is as given in the legend for Figure 8.}
\label{fig:fig9}
\end{figure}

In Figs. 5-6 we present our results for
$\eta$ electroproduction in comparison with experimental data.
In Fig. 5 the results are presented for differential
cross sections at $Q^2=0.375~(GeV/c)^2$ and $W=1.53~GeV$.
The data have some forward-backward
asymmetry which has not been seen in photoproduction
data and is not reproduced by our present analysis.
In Fig. 6 the data at $Q^2=0.75~(GeV/c)^2$ are presented
in the form of coefficients $A,~B,~C,~D,~E,~F$ 
in the expansion of the differential cross section over $cos\theta$, 
where  $\theta$ is the polar angle \cite{10}. 
The agreement with the data
is good everywhere, except for $B$ below $W=1.65~GeV$.
The coefficient $B$ and the forward-backward
asymmetry in the cross section come from interference between
$S_{11}$ and $P_{11}$ waves,
and the analysis shows that the contribution
$\simeq 0.2~\mu b^{1/2}$ is required in the
$ImM_{1-}$ in the $S_{11}(1535)$ resonance region
for both values of $Q^2=0.375~$ and $0.75~(GeV/c)^2$.
Such contribution is excluded by photoproduction
data. 
Our estimations show that the contribution
$\simeq 0.2~\mu b^{1/2}$ to $ImM_{1-}$ can not be explained  
by  $\gamma^*p\rightarrow \pi N\rightarrow \eta p$
and $\gamma^*p\rightarrow \eta p\rightarrow \eta p$, i.e.,
by final state interactions effects.
Due to its large width, the tail of the  $P_{11}(1440)$ can, in principle,
contribute to the $ImM_{1-}$ above the $\eta p$ production threshold. 
However, in this case one needs an unrealistically large
coupling constant $F_2^R(Q^2)$ 
for the $\gamma^* N\rightarrow P_{11}(1440)$
transition (notations of Ref. \cite{14}), 
by an order of magnitude larger than in 
$\pi$ electroproduction and  in
$\eta$ photoproduction \cite{14}.
Therefore, there is no common explanation 
of the forward-backward asymmetry in the electroproduction
data, and the consistency of photo-
and electroproduction data should be checked.
\section{Summary}
We present in this paper the results of our analysis of recent
CLAS data on $\pi$ and $\eta$ electroproduction \cite{8,9,10,11,12}.
High precision of the data in combination with their full angular coverage,
the existence of longitudinally polarized beam asymmetry data and utilization of two approaches,
allowed us to obtain accurate results for the photocoupling transition amplitudes
of the $P_{33}(1232)$, $P_{11}(1440)$, $D_{13}(1520)$,
and $S_{11}(1535)$ electroexcitation on the proton.

\begin{itemize}

\item{For the first time definite results are obtained
for the longitudinal amplitudes of the transitions
$\gamma^*p\rightarrow P_{11}(1440)$, $D_{13}(1520)$, and $S_{11}(1535)$.}

\item{The values obtained for the ratios $ImE_{1+}^{3/2}/ImM_{1+}^{3/2}$,
$ImS_{1+}^{3/2}/ImM_{1+}^{3/2}$ for the 
$P_{33}(1232)$ contribution at the resonance position
are in good agreement with the results
of Ref. \cite{8} obtained from $\pi^0$ cross sections
using a truncated multipole expansion.}

\item{We have found that the value of the neutron electric form factor 
is very important for the simultaneous description of the $\pi^+$
and $\pi^0$ cross sections in the $P_{33}(1232)$ resonance 
region. The obtained values are in good agreement
with the results of recent direct measurements \cite{22,23,24,25}.}

\item{The photocoupling amplitudes for the $\gamma^*p\rightarrow S_{11}(1535)$ transition
extracted separately from the analysis of $\pi$ and $\eta$ electroproduction are in
good agreement.}

\item{The transverse amplitude of the  $\gamma^*p\rightarrow P_{11}(1440)$
transition
falls very rapidly with increasing $Q^2$, and, apparently, changes
sign at $Q^2\simeq 0.5-0.6~(GeV/c)^2$.}

\item{Our results show a strong longitudinal
response for the  $P_{11}(1440)$; thus, they rule out 
presentation of this resonance as a $q^3G$ hybrid state.}

\item{Absolute values of the transverse amplitudes ${}_pA_{1/2}$ and 
${}_pA_{3/2}$ of the  
$\gamma^*p\rightarrow D_{13}(1520)$ transition become equal to each
other at $Q^2=0.4-0.65~(GeV/c)^2$.}

\item{Comparison of our results with
the existing quark model predictions show that none 
simultaneously gives a good description
of the $\gamma^*p\rightarrow   
P_{11}(1440)$, $D_{13}(1520)$, $S_{11}(1535)$
transition amplitudes.}

\item{In the analysis of $\eta$ electroproduction
in the $S_{11}(1535)$ resonance region, we have found it necessary
to introduce a large $P_{11}$ contribution
which is not seen in photoproduction data and is not related
to the $P_{11}(1440)$ resonance.} 

\end{itemize}

We are grateful to D. Richards for 
discussions of lattice results,
to V. Mokeev for useful discussions, to 
I. Strakovski for his help in getting data from SAID database, 
and to M. M. Giannini for providing us the results
of their quark model predictions.
I. G. Aznauryan expresses her gratitude for the hospitality 
at Jefferson Lab.

\widetext

\begin{table}
\begin{center}
\begin{tabular}{cccccc}
\hline
$Q^2~(GeV/c)^2$&$ImM_{1+}^{3/2}~(\mu 
b^{1/2})$&$ImE_{1+}^{3/2}/ImM_{1+}^{3/2}~(\%)$& 
$ImS_{1+}^{3/2}/ImM_{1+}^{3/2}~(\%)$&Approach\\
\hline
$0.4$&$4.93\pm 0.01$&$-2.4\pm 0.2$&$-5.0\pm 0.2$&IM\\
${}$&$4.97\pm 0.01$&$-2.9\pm 0.2$&$-5.9\pm 0.2$&DR\\
${}$&${}$&$-3.4\pm 0.4\pm 0.4$&$-5.6\pm 0.4\pm 0.6$&\cite{8}\\
\hline
$0.65$&$3.87\pm 0.01$&$-1.0\pm 0.3$&$-6.2\pm 0.4$&IM\\
${}$&$3.89\pm 0.01$&$-2.0\pm 0.3$&$-7.0\pm 0.4$&DR\\
${}$&${}$&$-1.9\pm 0.5\pm 0.5$&$-6.9\pm 0.6\pm 0.5$&\cite{8}\\
${}$&${}$&$-2.0\pm 0.4\pm 0.4$&$-6.6\pm 0.4\pm 0.2$&\cite{8}\\
\hline
\end{tabular}
\caption{\label{tab2} The results for the imaginary
parts of $M_{1+}^{3/2}$, $E_{1+}^{3/2}$, $S_{1+}^{3/2}$
at $W=1.229~GeV$. The results
on the third and fourth rows are obtained in Ref. \cite{8}
using a truncated multipole expansion. }  
\end{center}
\end{table}

\begin{table}
\begin{center}
\begin{tabular}{ccccccc}
\hline
${}$&$Q^2$&${}_pE_{0+}^{1/2}$&${}_pS_{0+}^{1/2}$&
${}_pA_{1/2}$&${}_pS_{1/2}$&Approach\\
${}$&$[(GeV/c)^2]$&$(\mu b^{1/2})$&$(\mu b^{1/2})$&
$(10^{-3}GeV^{-1/2})$&$(10^{-3}GeV^{-1/2})$&\\
\hline
$\gamma^* p\rightarrow \pi N$ &$0.4$&$0.56\pm 0.01$&$-0.20\pm 
0.02$&$95\pm 2$&$-25\pm 2$&IM\\
&${}$&$0.52\pm 0.02$&$-0.15\pm 0.02$&$88\pm 4$&$-18\pm 2$&DR\\
&$0.65$&$0.58\pm 0.02$&$-0.14\pm 0.02$&$98\pm 4$&$-17\pm 2$&IM\\
&${}$&$0.55\pm 0.02$&$-0.1\pm 0.02$&$93\pm 4$&$-12\pm 2$&DR\\
\hline
$\gamma^* p\rightarrow \eta p$ &$0.375$&$1.79\pm 0.02$&$-0.33\pm 
0.07$&$92\pm 1$&$-12\pm 3$&IM\\
&${}$&$1.77\pm 0.02$&$-0.36\pm 0.08$&$91\pm 1$&$-13\pm 3$&DR\\
&$0.75$&$1.73\pm 0.02$&$-0.32\pm 0.08$&$89\pm 1$&$-12\pm 3$&IM\\
&${}$&$1.80\pm 0.02$&$-0.38\pm 0.08$&$93\pm 1$&$-14\pm 3$&DR\\
\hline
\end{tabular}
\caption{\label{tab3} The $S_{11}(1535)$ amplitudes 
obtained in the analyses of $\pi$ and $\eta$ electroproduction.
Helicity amplitudes for the $S_{11}(1535)$
electroexcitation are obtained using $M=1530~MeV$, 
$\Gamma_{tot}=150~MeV$, $\beta_{\pi N}=0.4$,
$\beta_{\eta N}=0.55$.}  
\end{center}
\end{table}

\begin{table}
\begin{center}
\begin{tabular}{ccccccccc}
\hline
Resonance&$Q^2$&${}_pM_{l-}^{1/2}$&${}_pE_{l-}^{1/2}$&
${}_pS_{l-}^{1/2}$&${}_pA_{1/2}$&${}_pA_{3/2}$&${}_pS_{1/2}$&Approach\\
\hline
$P_{11}(1440)$&0.4&$0.07\pm 0.01$&${}$&$0.31\pm 0.01$&
$-15\pm 2$&&$44\pm 2$&IM\\
&&$0.02\pm 0.01$&${}$&$0.29\pm 0.01$&$-4\pm 2$&&$41\pm 2$&DR\\
&0.65&$-0.02\pm 0.02$&${}$&$0.31\pm 0.02$&$4\pm 4$&&$44\pm 4$&IM\\
&&$-0.11\pm 0.02$&${}$&$0.26\pm 0.02$&$23\pm 4$&&$37\pm 4$&DR\\
\hline
$D_{13}(1520)$&0.4&$0.28\pm 0.01$&$0.15\pm 0.01$&$-0.17\pm 0.01$&
$-66\pm 3$&$71\pm 4$&$-46\pm 3$&IM\\
&&$0.29\pm 0.01$&$0.16\pm 0.01$&$-0.15\pm 0.01$&$-68\pm 3$&$75\pm 
3$&$-41\pm 3$&DR\\
${}$&0.65&$0.27\pm 0.01$&$0.11\pm 0.01$&$-0.14\pm 0.01$&
$-67\pm 3$&$62\pm 4$&$-38\pm 3$&IM\\
&&$0.28\pm 0.01$&$0.12\pm 0.01$&$-0.14\pm 0.01$&$-69\pm 3$&$66\pm 
4$&$-38\pm 3$&DR\\
\hline
\end{tabular}
\caption{\label{tab4} Amplitudes of the $P_{11}(1440)$
and $D_{13}(1520)$ obtained in the analysis of $\pi$
electroproduction. Multipole amplitudes are in $\mu b^{1/2}$
units, helicity amplitudes are in $10^{-3}GeV^{-1/2}$ units.
Helicity amplitudes for the $P_{11}(1440)$ electroexcitation
are obtained using $M=1440~MeV$,
$\Gamma_{tot}=350~MeV$, $\beta_{\pi N}=0.6$.
Helicity amplitudes for the $D_{13}(1520)$ electroexcitation
are obtained using $M=1520~MeV$,
$\Gamma_{tot}=120~MeV$, $\beta_{\pi N}=0.5$.}
\end{center}
\end{table}

\narrowtext

\end{document}